\documentclass{article}
\usepackage{spconf,amsmath,graphicx}
\usepackage{multirow}
\usepackage{color}
\usepackage{enumitem}
\usepackage{caption}
\usepackage{subcaption}
\usepackage{url}

\newcommand{\comment}[1]{}

\usepackage[linesnumbered,ruled,noresetcount]{algorithm2e}
\SetAlFnt{\small} %{\footnotesize} % Small algorithm code.
\SetKwBlock{KwMain}{main}{}
\SetKwBlock{KwInit}{initialize}{}
\SetKwBlock{KwFunction}{function}{}
\SetKwBlock{Begin}{begin}{end} 
\SetKwFunction{KwFuncSort}{Sort}

\newcommand{\TRUE}{\ensuremath{ \textsc{true} }}
\newcommand{\FALSE}{\ensuremath{ \textsc{false} }}

\newcommand{\funcSum}{\ensuremath{ \texttt{GroupComputeSum} }} 
\newcommand{\funcTopK}{\ensuremath{ \texttt{DATA} }} 
\newcommand{\funcThresh}{\ensuremath{ \texttt{GroupComputeThreshold} }} 
\newcommand{\genX}{\ensuremath{ \texttt{GenerateVectorFromList} }}

\newcommand{\mList}[1]{\ensuremath{L^{#1}}}

\newcommand{\mb}[1]{\ensuremath{b^{#1}}}
\newcommand{\mA}[1]{\ensuremath{A^{#1}}}
\newcommand{\my}[2]{\ensuremath{y^{#1}_{#2}}}
\newcommand{\mz}[2]{\ensuremath{z^{#1}_{#2}}}
\newcommand{\me}[1]{\ensuremath{\epsilon^{#1}}}
\newcommand{\xh}[1]{\ensuremath{x_{#1}}}
\newcommand{\mx}[2]{\ensuremath{x^{#1}_{#2}}}

\newcommand{\thresh}[1]{\ensuremath{{\mathcal{T}}_{_K}\left(#1\right)}}
\newcommand{\tp}{\ensuremath{^{\mathsf{T}}}}
\newcommand{\xopt}{\ensuremath{\hat{x}}}

\newcommand{\oid}{\ensuremath{o}}
\newcommand{\ov}[1]{\ensuremath{val^{#1}(\oid)}}
\newcommand{\gtop}[1]{\ensuremath{\overline{\tau}^{#1}}}
\newcommand{\gbot}[1]{\ensuremath{\underline{\tau}^{#1}}}
\newcommand{\topThresh}{\ensuremath{\gtop{p}}}
\newcommand{\bottomThresh}{\ensuremath{\gbot{p}}}

\newcommand{\dihtT}{\ensuremath{T_1}}
\newcommand{\bpT}{\ensuremath{T_2}}
\title{Distributed Sparse Signal Recovery For Sensor Networks}
\name{Stacy Patterson, Yonina C. Eldar, and Idit Keidar\thanks{The work of Y. Eldar is supported in part by the Israel Science Foundation under Grant no.
170/10, and in part by the Ollendorf Foundation.  The work of S. Patterson is funded in part by the Arlene \& Arnold Goldstein Center at the Technion Autonomous Systems Program, 
    a Technion fellowship, and an Andrew and Erna Finci Viterbi Fellowship.}}
\address{Department of Electrical Engineering \\Technion - Israel Institute of Technology, Haifa, Israel\\
\{stacyp,yonina,idish\}@ee.technion.ac.il}

\begin{document}
%\ninept
%
\maketitle
\begin{abstract}
We propose a distributed algorithm for sparse signal recovery in sensor networks based on Iterative Hard Thresholding (IHT). Every agent has a set of measurements of a signal $x$, and the objective is for the agents to recover $x$ from their collective measurements at a minimal communication cost and with low computational complexity. A na\"{i}ve distributed implementation of IHT would require global communication of every agent's full state in each iteration. We find that we can dramatically reduce this communication cost by leveraging solutions to the distributed top-$K$ problem in the database literature. Evaluations show that our algorithm requires up to three orders of magnitude less total bandwidth than the best-known distributed basis pursuit method. \end{abstract}
\vspace{-.2cm}
\section{Introduction}
\vspace{-.2cm}
\label{sec:intro}

%%%% Basic idea of compressed sensing%
In compressed sensing, a sparse signal $x \in \mathbf{R}^N$ is sampled and compressed into a set of $M$ measurements, where $M$
is typically much smaller than $N$.  If these measurements are taken appropriately, then it is possible to recover $x$
from  this small set of  measurements \cite{DE11}. %$K \log(N)$

Compressed sensing is an appealing approach for sensor networks, where measurement capabilities may be limited 
due to both coverage and energy constraints. Recent works have demonstrated that compressed sensing is applicable to a variety of 
sensor networks problems including event detection \cite{MLH09}, urban environment monitoring \cite{LZZL11} and traffic estimation \cite{YZZ10}. 
In these applications, measurements of the signal are taken by sensors that are distributed throughout a region.
The measurements are then collected at a single fusion center where signal recovery is performed.  
%Due to limits in bandwidth, storage, and computation capabilities, such centralized schemes may not always be feasible.
Due to limits in bandwidth, storage, and computation capabilities, it may be more efficient, and sometimes even necessary, to perform signal recovery 
in the network in a distributed fashion.
% Need another sentence here about why we need distributed solution.

Distributed solutions for compressed sensing have begun to receive attention lately.  
For example, one work proposes a distributed subspace pursuit recovery algorithm for a mixed-support set model \cite{SCS12}.
This work assumes that every agent knows the sensing matrix of every other agent. The need for global knowledge of these matrices presents a scalability bottleneck
as individual sensors do not have the capacity to store and process a large number of these matrices.
Several works have proposed distributed basis pursuit algorithms for sparse signal recovery in sensor networks where the measurement matrices are not globally known \cite{BG10,MXAP11,MXAP12}.
In these algorithms, agents collaborate to solve a convex relaxation of the original recovery problem. 
Each agent stores its own estimate of the signal $x$, and, in each iteration, it updates this estimate based on communication with its neighbors in the network.
This approach requires that every agent solve a local convex optimization problem in each iteration.
While these algorithms use only local communication, each agent must send its  entire estimate vector to every neighbor in every iteration.  
This vector is not necessarily sparse until the algorithm converges, and therefore, the messages can be quite large.  
As a result, these algorithms have a large total bandwidth cost.
Furthermore, simulations show that this bandwidth cost increases dramatically as the network connectivity increases.  

We propose an alternative approach to distributed sparse signal recovery in sensor networks that is based on \emph{Iterative Hard Thresholding} (IHT) \cite{BD09}.
%n centralized settings, greedy solutions like IHT have computational benefits over methods like basis pursuit that are based on a convex relaxation. 
%Same benefits in D-IHT over distributed basis pursuit
In our distributed implementation of IHT, which we call D-IHT, all agents store identical copies of the estimate  of $x$.  
In each iteration, every agent first performs a \emph{local computation} 
to derive an intermediate vector.  The agents then perform a \emph{global computation} on their intermediate vectors to derive the next iterate.
A na\"{i}ve distributed implementation of IHT would require global communication of the intermediate vector of each agent  in every iteration. 
We find that we can dramatically reduced the communication cost of this global computation by leveraging solutions to the distributed top-$K$ problem in the database literature \cite{NR99,FLN03,IAE04}.
Our evaluations show that D-IHT requires up to  three orders of magnitude less total bandwidth than the best-known  distributed basis pursuit method.
D-IHT is also computationally simpler since it does not require that agents solve local convex optimization problems.
While, in this work, we present our distributed recovery algorithm for compressed sensing, we note that our solution easily generalizes to sparse signal recovery from 
nonlinear measurements \cite{BE12}.

%We believe our technique will also prove useful in developing distributed implementations of other sparse recovery algorithms. 

The remainder of the paper is organized as follows.  In Section~\ref{problem.sec}, we detail our problem setting and formulation and provide a brief description of IHT.
The D-IHT algorithm is presented in Section~\ref{algorithm.sec}.  
Section~\ref{sec:results} gives numerical results on the performance of D-IHT. 
 %%%%%%%%%%%%%%%%%%%%%%%%%%%
%%%%%%%%%%%%%%%%%%%%%%%%%%%
%%%%%%%%%%%%%%%%%%%%%%%%%%%

\vspace{-.3cm}
\section{Preliminaries}
\label{problem.sec}
\vspace{-.2cm}
\subsection{Problem Formulation} 
\vspace{-.2cm}
We consider a set of $P$ agents that form a connected, undirected static network topology with $E$ edges.  The agents may be the sensors themselves or they may be fusion nodes that collect measurements from several nearby sensors.  Every agent knows the number of agents $P$, and we assume there is a unique agent identified as agent 1. 
If the uniquely identified agent is not defined a priori, one can be chosen using a variety of well-known distributed algorithms (see \cite{L96}).
Agents communicate with their neighbors in the network using fixed size messages.  Messaging is reliable but asynchronous, meaning that every message that is sent is eventually delivered, but the delay between sending and delivery may be arbitrarily long. 

There is a $K$-sparse signal  $x \in \mathbf{R}^N$ that the agents seek to estimate. 
Each agent $p =1 \ldots P$ has $M_p > 0$ (possibly noisy) measurements of $x$ that have been taken using the agent's sensing matrix $\mA{p} \in \mathbf{R}^{M_p \times N}$.
There are $M = M_1 + \ldots + M_P$ measurements in total.
The measurement vector of agent $p$, denoted $\mb{p}$, is given by $\mb{p} = \mA{p} x + \me{p}$,
where $\me{p} \in \mathbf{R}^{M_p}$ is the measurement error for agent $p$.
Agents do not know the sensing matrices or measurement vectors of other agents.

Our goal is for every agent to recover the same signal $x$ from their collective measurements at a minimal communication cost.
%We now formalize the recovery problem.
Let $b$ be the vector of all measurements, and let $A$ be the sensing matrix for the entire system:
\[
\renewcommand*{\arraystretch}{1.2}
b := \left[ \begin{array}{c}
\mb{1} \\
\hline
\vdots \\
\hline
\mb{P} 
\end{array} \right], ~~~~~~~~~~~~~ A :=  \left[ \begin{array}{c}
~~~~~~~~\mA{1}~~~~~~~~~ \\
\hline 
\vdots \\
\hline
\mA{P}
\end{array} \right].
\] 
To recover $x$ from $A$ and $b$, the agents must solve the following optimization problem,
\begin{equation}
 \xopt  = \arg \min_{x \in \mathbf{R}^N} \| A x - b\|_2^2~~~\mbox{subject to}~~\|x\|_0 \leq K,
 \label{opt.eq}
\end{equation}
where $\| \cdot \|_0$ denotes the $l_0$ norm, i.e, the number of non-zero components.

This problem is known to be NP-Hard in general \cite{N95}.
However, for suitable $A$ matrices, efficient centralized algorithms to recover $\xopt$ exist.
Our distributed solution is based on IHT \cite{BD08,BD09}, which we describe next.

\subsection{Iterative Hard Thresholding Algorithm}

IHT  is a gradient-like, iterative algorithm for finding a $K$-sparse vector $\xopt$ in a centralized setting where $A$ and $b$ are known.  
Let  $\thresh{v}$ be the thresholding operator which returns a vector where all but the $K$ entries of $v$ with the largest magnitude are set to 0 (with ties broken arbitrarily).	
The IHT algorithm begins with an initial, arbitrary $K$-sparse vector $\xh{0}$.
In each iteration, a gradient-step is performed, followed by  application of the thresholding operator:
\label{sec:algorithm}
\begin{equation}
\xh{t+1}= \thresh{\xh{t} - \alpha A\tp \left(b - A\xh{t}\right)}.
\label{iht.eq}
\end{equation}
It has been shown that, for $\alpha < 1/\left(2 \lambda_{max}(A\tp A) \right)$,  IHT converges to a local minimum of (\ref{opt.eq}) \cite{BD09,BE12}.
%\note{Has linear convergence rate been proven for generalized $\alpha$?  \cite{BD09} proves it for $\alpha = 1$ so long as $\|A_2\| < 1$}

We note that, even if $A$ satisfies the properties necessary to enable recovery using IHT,
it is not necessary and, in fact, not likely that each $\mA{p} $ satisfies these properties.
Therefore it is not possible for any single agent to recover $\xopt$ on its own; 
agents must exchange information with one another to perform the recovery.
In the next section, we present D-IHT, our distributed implementation of IHT.

%%%%%%%%%%%%%%%%%%%%%%%%%%%
%%%%%%%%%%%%%%%%%%%%%%%%%%%
%%%%%%%%%%%%%%%%%%%%%%%%%%%

\begin{figure*}[htb]
\smaller
\begin{subfigure}{.46\linewidth}
\centering
\def\arraystretch{1.1}%
\begin{tabular}{|c|c||c|c||c|c|}
\hline
\multicolumn{2}{|c||}{\textbf{Agent 1}} & \multicolumn{2}{|c||}{\textbf{Agent 2}} & \multicolumn{2}{|c|}{\textbf{Agent 3}} \\
\hline
\textbf{$\mz{1}{t}$} & \textbf{$L^1$} & \textbf{$\mz{2}{t}$} & \textbf{$L^2$} & \textbf{$\mz{3}{t}$} & \textbf{$L^3$} \\
\hline
\multirow{10}{*}{$\left[\begin{array}{c}
21 \\
14 \\
11 \\
13 \\
2 \\
4 \\
10 \\
6 \\
12 \\
1
\end{array}\right]$} & $(1,21)$ & \multirow{10}{*}{$\left[\begin{array}{c}
28 \\ 
3 \\
26 \\
45 \\
20 \\
10 \\
1 \\
13 \\
18 \\
22
\end{array}\right]$} & $(4,45)$ & \multirow{10}{*}{$\left[\begin{array}{c}
2 \\
5 \\
30 \\
14 \\
6 \\
15 \\
27 \\
1 \\
29 \\
7 
\end{array}\right]$}  & $(3,30)$ \\
& $(2,14)$ & & $(1,28)$ &  & $(9,29)$ \\
 & $(4,13)$ &  & $(3,26)$ &  & $(7,27)$ \\
 & $(9,12)$ &  & $(10,22)$ &  & $(6,15)$ \\
 & $(3,11)$ &  & $(5,20)$ &  & $(4,14)$ \\
 & $(7,10)$ &  & $(9,18)$ &  & $(10,7)$ \\
& $(8,6)$ &  & $(8,13)$ &  & $(5,6)$ \\
& $(6,4)$ &  & $(6,10)$ &  & $(2,5)$  \\
& $(5,2)$ &  & $(2,3)$ &  & $(1,2)$ \\
& $(10,1)$ &  & $(7,1)$ &  & $(8,1)$ \\
\hline
\end{tabular}
\caption{The vector $\mz{p}{t}$ and the resulting sorted list $L^p$, at three agents.}
\vspace{.15cm}
\label{scores.fig}
\end{subfigure}
\hspace{.4cm}
\begin{subfigure}{.51\linewidth}
\centering
\def\arraystretch{1.2}%
\begin{tabular}{|c|c|c|c|c|c|c|c|c|}
\hline
\textbf{step} & \textbf{agent} & \textbf{object} & \textbf{sum} & $\tau_1$ & $\tau_2$ & $\tau_3$ & $\tau$ & \textbf{top-$2$ set}\\
\hline
1 & 1 & 1 & 51 & 21 & ? & ? & ?  & $\{(1,51)\}$\\
2 & 2 &  4 & 72 & - & 45 & ? & ? & $\{(4,72), (1,51)\}$\\
3 & 3 & 3 & 67 & - & -  & 30 & 96 & $\{(4,72), (3,67)\}$ \\
4 & 1 & 2 & 22 & 14 & - & - & 89 & $\{(4,72), (3,67)\}$ \\
5 & 2 & 10 & 30 & - & 22 & - & 66 & $\{(4,72), (3,67)\}$ \\ 
\hline
\end{tabular}
\caption{Steps of TA to find top two objects.  After five steps,  the threshold $\tau$ is 65, and objects 3 and 4 both have sums greater than $\tau$.  No remaining objects can have a sum
greater than $\tau$.  Therefore, the top two objects have been found, and the algorithm terminates.}
\label{TAsteps.fig}
\end{subfigure}
\caption{Example execution of the TA algorithm for $K=2$.} \label{TA.fig}
\end{figure*}

\section{Distributed Iterative Hard Thresholding}
\label{algorithm.sec}

In D-IHT, every agent stores an identical copy of $\xh{t}$, which is initially 0.
In each iteration $t$, each agent first performs a \emph{local computation} 
to derive an intermediate vector $\mz{p}{t} \in \mathbf{R}^N$.  The agents then perform a \emph{global computation} on their intermediate vectors to derive the next iterate
$\xh{t+1}$, which is, again, identical at every agent.
We now define these local and global computations.

\vspace{.1cm}
\noindent \textbf{Local computation.}  Each agent $p$ computes a local residual vector, $\my{p}{t} := \mb{p} - \mA{p} \xh{t}$. 
The intermediate vector for agent $p$, denoted $\mz{p}{t}$, is then computed as follows,
\begin{equation}
\mz{p}{t} = \left\{ \begin{array}{ll}
 \xh{t} - \alpha \left(\mA{p}\right)\tp \my{p}{t} & \mbox{if}~~p=1, \\
- \alpha \left(\mA{p}\right)\tp \my{p}{t}  & \mbox{otherwise}.
  \end{array} \right. \label{eq:z}
\end{equation}
Note that each agent can compute $\mz{p}{t}$ using its local information.
%We assume that there is one node with ID = 1.  Can be selected as a pre-processing step using XXX messages.

\vspace{.1cm}
\noindent \textbf{Global computation.}
In the global computation step, all agents must compute a function $G$ that depends on all of their intermediate vectors.  This function is 
defined as follows,
\begin{equation}
\xh{t+1} = G\left(\mz{1}{t}, \ldots, \mz{P}{t}\right) := \thresh{\sum_{p=1}^P \mz{p}{t}}. \label{global.eq}
\end{equation}
We note that the combination of the local computation step (\ref{eq:z}) and the global computation step (\ref{global.eq}) are equivalent to (\ref{iht.eq}).

A na\"{i}ve implementation of $G$ is for all agents to collaborate to compute all $N$ sums, one for each component of the intermediate vectors.
Then, each agent can independently determine the values with the $K$ largest magnitudes.
This approach requires communication of all components of all intermediate vectors and is thus very costly with respect to bandwidth.
We derive a more communication efficient approach by leveraging work in the database literature on the distributed top-$K$ problem.
We describe this problem and a popular solution in Section \ref{topk.sec}.  We then present our distributed algorithm for computing $G$ in Section \ref{distabs.sec}.

\vspace{-.2cm}
\subsection{The Distributed Top-$K$ Problem}
\label{topk.sec}

In the distributed top-$K$ problem, each agent $p$ has a list $L^p$ of pairs $(\oid, \ov{p})$,
 where $\oid$ is an object ID and $\ov{p}$ is the value of object $\oid$ at agent $p$.
 In our recovery problem, this list is generated from the vector $\mz{p}{t}$; $\oid$ is the index into the vector $\mz{p}{t}$ and
$\ov{p}$ is the value of $\mz{p}{t}$ at index $\oid$.
Each object has a score; in our case, the score is the magnitude of the sum of object's values at all agents.
The objective is to find the objects with the $K$ largest scores.
Clearly, it is possible to solve this problem by computing the sum for every object and then selecting the objects with the $K$ largest magnitude sums,
but it is often not necessary to compute all sums in order to find the top-$K$ objects. 

The \emph{Threshold Algorithm} (TA) is a solution for the distributed top-$K$ problem that is instance optimal, i.e., TA
makes the minimum number of sum computations necessary for a given input of lists \cite{NR99,FLN03,IAE04}.  
As it was originally proposed,
TA requires that all values be non-negative\footnote{More precisely, TA requires that that an object's score is given by a function that is monotonic in the values.}.  
Here, we present the algorithm, assuming that this is the case.  In Section \ref{distabs.sec}, we explain how we modify TA to support both negative and non-negative values.

In TA, each agent first sorts its list by object value, in descending order.  
One agent acts as the leader, requesting information from the other agents, computing sums for objects, and distributing the final top-$K$ list to all agents.  
The algorithm proceeds as follows.
The leader requests an object/score pair from each agent's list in sorted order, one pair from one agent in any given step.  When the leader receives a pair containing an object it has not yet seen, it requests the value for that object from all other agents and computes the object's sum.
The leader always stores the objects with the $K$ largest sums it has seen so far.
In addition, the leader stores the value of the last object seen from each agent $p$ under the sorted access.  This value is denoted $\tau_p$. 
It computes the threshold value $\tau = \tau_1 +  ... \tau_P$ in each step.
As soon as the leader has seen $K$ objects that each have a score of at least $\tau$, the algorithm terminates. The leader then disseminates the list of top-$K$ objects and their scores.

An example execution of TA is given in Figure \ref{TA.fig}.
Note that, while each list has 10 objects, the algorithm only requires five sum computations to find the top two objects.

% We denote $i^th$ element in agent $p$'s list by $L_i^p = (L_i^p.id, L_i^p.val)$, 

\begin{algorithm}[t]
\caption{Pseudocode for D-IHT algorithm.}
\label{diht.alg}
\small
\SetAlgoNoLine
\SetAlgoNoEnd
\DontPrintSemicolon
\SetNoFillComment
\KwInit{
$\mx{p}{0} \gets 0$ \;
$t \gets 0$ \;
}
\While{$\TRUE$}{
$\mz{p}{t} \gets$ value from equation (\ref{eq:z})  $~~~~~~~~~$\hfill \emph{Local computation.}\;
$\mList{p} \gets \KwFuncSort(\mz{p}{t})~~~~~~~~~~~~~~~~~~~$ \hfill \emph{Create sorted list from $\mz{p}{t}$.} \;
$\mx{p}{t+1} \gets \funcTopK(\mList{p})~~~~~~~~~~~~~~~~~~~~~~~~$ \hfill \emph{Global computation.} \;
 $t \gets t+1$ \;
}
\vspace{-.1cm}
\end{algorithm}

\subsection{Distributed Computation of $G$} \label{distabs.sec}
\vspace{-.1cm}
The computation of $G$ is equivalent to solving a top-$K$ problem over the vectors $\mz{p}{t}$, $p=1 \ldots P$.
In D-IHT, we solve this top-$K$ problem using a modified version of TA that is not leader-based and that accommodates both negative and non-negative values.
We describe our modifications and the resulting algorithm below.

\vspace{.15cm}
\noindent \textbf{Support for non-negative values.} 
As it was originally proposed, TA is applicable only to score functions like sum that are monotonic in the object values.  
 To compute $G$, we need to find the top-$K$ \emph{magnitude} sums, which means that the score function is not monotonic unless all values are non-negative.  
 A simple way to address this limitation is to run two instances of TA to find the top-$K$ largest sums and the top-$K$ smallest sums (since a sum with a negative value may have a large magnitude).
 The top-$K$ magnitude objects can then be found from this set of $2K$ objects.
 We implement a more efficient algorithm in which the agents find the top-$K$ magnitude sums in a single algorithm instance by processing the sorted lists from both the top and the bottom.
 
\vspace{.15cm}
\noindent \textbf{Decentralized list processing.}
We speed up the execution of  TA by having each agent independently processes its own list in sorted order. Each agent initiates a \emph{group sum computation} for each new object it encounters, so that $P$ group sum computations are executed in parallel for each iteration of the algorithm. 
In a group sum computation, every agent learns the sum of the values for the specified object. There are many distributed algorithms for group sum computation (e.g. \cite{MFHH02,KDG03}). We use an algorithm based on the well-known \emph{broadcast-convergecast} paradigm (see \cite{S83}). 
A request message is propagated down a broadcast tree that is rooted at the initiating agent.  Each agent collects values from its children (if it has any), sums up these values with its own, and sends the result to its parent. For a network with $E$ edges, the algorithm requires a preprocessing phase of less than $2E$ messages to create a broadcast tree. 
Each group sum computation requires  $3(P-1)$ messages,  and each agent $p$ sends at most $3D_p$ messages, where $D_p$ is the node degree.

\vspace{.15cm}
\noindent \textbf{The algorithm.}
We call our modified version of TA the Distributed Absolute Threshold Algorithm (DATA).
We briefly summarize the algorithm below. The pseudocode is given in the appendix.

In DATA, Each agent stores two variables for global thresholds, one for sorted access from the top of the lists, denoted $\gtop{p}$,  and one for sorted access from the 
bottom of the lists, denoted $\gbot{p}$.  Both values are initially $\infty$.
Each agent also stores a top-$K$ list that is initially empty.
The agents each execute the following steps.
\begin{enumerate}[leftmargin=0cm,itemindent=.35cm,labelwidth=\itemindent,labelsep=0cm,align=left,itemsep=-.1cm]
\item If $\gtop{p} > \gbot{p}$, select next object from top of list for which $p$ has not received a sum in a previous iteration.  
Else, select next object from bottom of list for which $p$ has not received a sum in a previous iteration
\item Initiate group sum computation for selected object.  Agent 1 also initiates a group sum computation for the global threshold, either $\gtop{p}$ or $\gbot{p}$, depending
on whether it accessed its object from the top or bottom of its list.  
When participating in the group sum computation for $\gtop{p}$, an agent uses the most recent value seen in sorted access from the top of its list,
and for  $\gbot{p}$, it uses the most recent value seen in sorted access from the bottom.
\item On receipt of sums (from group computation) from all agents, update top-$K$ list with objects that have sums with the $K$ largest magnitudes seen so far. 
\item On receipt of threshold (from group computation), update appropriate global threshold variable, $\gtop{p}$ or $\gbot{p}$.
If the top-$K$ list contains $K$ objects with magnitudes at least $\max(|\gtop{p}|,|\gbot{p}|)$, return the list. Else, go to Step 1.
\end{enumerate}
We note that in a single iteration, multiple agents may initiate group sum computations for the same object.  While, in theory, this introduces unnecessary message overhead, in practice
the redundant sums do not add significantly to the total message cost.

\subsection{The D-IHT Algorithm and Analysis}

We combine the local computation step with DATA to arrive at the full D-IHT algorithm.
The pseudocode is given in Algorithm \ref{diht.alg}.
We note that while agent 1 plays a unique role in the local and global computations, it performs the same 
number and types of computations and sends the same number of messages as any other agent.  

\vspace{.15cm}
\noindent \textbf{Storage complexity.} Both D-IHT and distributed basis pursuit \cite{BG10,MXAP11,MXAP12} require $O(N)$ storage per agent.  

\vspace{.15cm}
\noindent \textbf{Message complexity.}
Let $\dihtT$ be the number of iterations of D-IHT required to achieve a certain accuracy.  
Let $S_j$ be the number of group sum computations for iteration $j$ (including the group threshold computations).
Each sum computation requires $3(P-1)$ messages.  Therefore, the total number of messages is $3(P-1) \sum_{j=1}^{\dihtT} S_j$.
We compare this to distributed basis pursuit where, in every iteration,  each agent sends its estimate of $x$ to all of its neighbors.
Assuming that the message size is limited to a single value, $N$ messages are required to send a single estimate.   
Let $\bpT$ be the number of iterations of distributed basis pursuit required to achieve the same accuracy as  $\dihtT$ iterations of D-IHT. The total number of messages sent in distributed basis pursuit is $2NE \bpT$.
For a connected network, $E  \geq P - 1$, and therefore, the total number of messages is at least $2N \bpT(P-1)$.

The preprocessing phase of D-IHT requires at most 2EP messages, which is less than the number of messages required for one iteration of distributed basis pursuit.
Therefore, if $\dihtT < \bpT$, then D-IHT requires fewer messages than distributed basis pursuit so long as less than $\frac{2}{3}N$ sums are computed per iteration of D-IHT, on average.
In our evaluations, $\dihtT$ is always at least one order of magnitude smaller than $\bpT$, and in most cases, the average number of sums computed per iteration of D-IHT is far fewer than $\frac{2}{3}N$.
%\note{In most cases, the average is less than XXX.}

%\noindent \textbf{Computational complexity.}
%With respect to computational complexity, in basis pursuit algorithms, every agent solves a local convex optimization problem in every iteration.
%In D-IHT, per iteration, agent needs to sort list, compute at most $D_p$ addition operations per sum, and perform comparison to update top-$K$ list.
%Is sorted less complicated than solving optimization problem.
%
	
%%%%%%%%%%%%%%%%%%%%%%%%%%%
%%%%%%%%%%%%%%%%%%%%%%%%%%%
%%%%%%%%%%%%%%%%%%%%%%%%%%%
\begin{table}
\caption{Recovery problem parameters.}  \label{params.tab}
\centering
\def\arraystretch{1.3}%
\smaller
\begin{tabular}{c|ccccc}
\textbf{Problem}& $N$ & $M$ & $P$ & $K$ & $\alpha$ \\
\hline 
Random  & 1000 & 250 & 50 & 20 & 1 \\
Sparco 7  & 2560 & 600 & 40  &  20 & 0.99 \\
Sparco 11 & 1024 & 256 & 64  &  32 & 0.0025 \\
Sparco 902 & 1000 & 200 & 50 & 3 & 0.99
\end{tabular}
\end{table}

\begin{table}
\caption{Evaluation results for D-IHT and D-ADMM.} \label{results.tab}
    \begin{subtable}[b]{1\columnwidth}
        \centering
\caption{ER graph with connection probability of 0.25.} \label{er25.tab}
\def\arraystretch{1.1}%
\smaller
\begin{tabular}{c|cc|cc}
\multirow{2}{*}{\textbf{Problem}} & \multicolumn{2}{c}{\textbf{Total Messages}} & \multicolumn{2}{c}{\textbf{Clock Ticks}}\\
& \textbf{D-IHT} & \textbf{D-ADMM} & \textbf{D-IHT} & \textbf{D-ADMM}  \\
\hline 
Random & $1.06 \times 10^6$ & $1.43\times 10^8$ & $5.13 \times 10^3$ & $1.60 \times 10^6$ \\
Sparco 7 & $2.23 \times 10^6$ & $1.11 \times 10^8$ & $2.59 \times 10^4$  & $2.02 \times 10^6$ \\
Sparco 11 & $3.28 \times 10^6$ & $5.25 \times 10^8$ & $1.24 \times 10^4$ & $4.29 \times 10^6$\\
Sparco 902 &$1.48 \times 10^6$ & $5.58 \times 10^7$ & $9.09 \times 10^3$ & $7.20 \times 10^5$
\end{tabular}
\vspace{.5cm}
\end{subtable}
\begin{subtable}[b]{1\columnwidth}
\centering
\caption{ER graph with connection probability of 0.75.} \label{er75.tab}
\def\arraystretch{1.2}%
\smaller
\begin{tabular}{c|cc|cc}
\multirow{2}{*}{\textbf{Problem}} & \multicolumn{2}{c}{\textbf{Total Messages}} & \multicolumn{2}{c}{\textbf{Clock Ticks}}\\
& \textbf{D-IHT} & \textbf{D-ADMM} & \textbf{D-IHT} & \textbf{D-ADMM}  \\
\hline 
Random & $1.13 \times 10^6$ & $1.21 \times 10^9$  & $3.75 \times 10^3$ & $1.18 \times 10^7$  \\
Sparco 7 & $2.27 \times 10^6$ & $7.33 \times 10^8$  & $9.66 \times 10^3$ & $9.33 \times 10^6$ \\
Sparco 11 & $3.41 \times 10^6$  & $4.88 \times 10^9$ & $8.06 \times 10^3$  & $3.41 \times 10^7$  \\
Sparco 902 &$1.54 \times 10^6$ & $2.67 \times 10^8$  & $5.78  \times 10^3$  & $2.65 \times 10^6$
\end{tabular}
\vspace{.5cm}
\end{subtable}
\begin{subtable}[b]{1\columnwidth}
\centering
\caption{Geometric graph with d=0.5.}\label{geo.tab}
\def\arraystretch{1.2}%
\smaller
\begin{tabular}{c|cc|cc}
\multirow{2}{*}{\textbf{Problem}} & \multicolumn{2}{c}{\textbf{Total Messages}} & \multicolumn{2}{c}{\textbf{Clock Ticks}}\\
& \textbf{D-IHT} & \textbf{D-ADMM} & \textbf{D-IHT} & \textbf{D-ADMM}  \\
\hline 
Random & $1.05 \times 10^6$ & $7.22 \times 10^7$  & $2.46 \times 10^4$ & $1.68 \times 10^6$   \\
Sparco 7 & $2.23 \times 10^6$ & $6.06 \times 10^7$  & $7.60 \times 10^4$ & $1.99 \times 10^6$ \\
Sparco 11 & $3.26 \times 10^6$  & $2.37 \times 10^8$ & $6.24 \times 10^4$  & $3.66 \times 10^6$  \\
Sparco 902 &$1.46 \times 10^6$ & $2.94 \times 10^7$  & $5.17 \times 10^4$  & $6.88 \times 10^5$
\end{tabular}
\end{subtable}
\end{table}
\section{Numerical Results}

\label{sec:results}
In this section, we present an experimental comparison of D-IHT and distributed basis pursuit.  As a representative example,
we select D-ADMM, a distributed implementation of the alternating direction method of multipliers that has been shown to outperform other distributed basis pursuit algorithms in
terms of the number of communications in similar experiments \cite{MXAP12}.  In  each iteration of D-ADMM, each agent exchanges its estimate with its neighbors and generates 
a new estimate by solving a local optimization problem involving its estimate and the estimates of some of its neighbors.
We have implemented D-IHT and D-ADMM in Matlab, using CVX \cite{GB11} to solve the local optimization problems in D-ADMM. 
D-ADMM requires a graph coloring, which we generate using the heuristic from the Matgraph toolbox \cite{S12}, as is done in \cite{MXAP12}. 
We include the preprocessing phase in our results for D-IHT, but we do not include graph coloring pre-processing in our results for D-ADMM.

\vspace{.1cm}
\noindent \textbf{Recovery problems.}
We evaluate the performance of D-IHT and D-ADMM on four reconstruction problems, similar to those in \cite{MXAP12}.  For the first problem, we generate the $A$ matrix with i.i.d Gaussian entries with zero mean and
variance of $1/m$.  The remaining three problems are from the Sparco toolbox \cite{BFHH07}.
The parameters for each problem are given in Table \ref{params.tab}.  For each problem, we divide the $A$ matrix evenly among the agents so that each agent has $M/P$
rows.  
For the randomly generated problem, we find the optimal sparse solution $\hat{x}$ using CVX.  For the Sparco problems, we use the provided optimal sparse solution.

\vspace{.1cm}
\noindent \textbf{Performance measures.}
For each algorithm, we measure the total number of messages sent in order for $\|\mx{p}{t} - \hat{x}\| / \|\hat{x}\| \leq 10^{-2}$ for all agents. 
To standardize the bandwidth comparison between the algorithms, we assume that only one value is sent per message.  Therefore, in D-ADMM, when an agent sends its $N$-vector
to its neighbor, this  requires $N$ messages.  D-IHT is designed so that only one value is sent per message. 
We also measure the time required for convergence in a synchronous network where each message is delivered in one clock tick.  For both algorithms, we only allow one message to be sent on a link in each direction per clock tick.

\vspace{.1cm}
\noindent \textbf{Results.}
Table \ref{results.tab} shows the results of our evaluations in three different network topologies. 
The first is an Erd\"{o}s-R\'{e}nyi (ER) graph \cite{ER59} where each pair of vertices is connected with probability 0.25 (Figure \ref{er25.tab}),
and the second is an ER graph where each pair of vertices is connected with probability 0.75 (Figure \ref{er75.tab}).
The third network topology is a geometric graph \cite{P04} with vertices placed uniformly at random in a unit square, and two vertices are connected if
they are within a distance of 0.5 of each other (Figure \ref{geo.tab}).

These results show that, for every recovery problem, D-IHT requires far fewer total messages than D-ADMM to achieve the same recovery accuracy,
between one and two orders of magnitude in most cases. D-IHT also requires less total time to perform the recovery than does D-ADMM.
We note that, as network connectivity increases, in D-ADMM the total message count and total time increase (Table \ref{er25.tab} vs. Table \ref{er75.tab}).  In D-IHT, sums can be computed more quickly in networks that are more connected.  Therefore, in D-IHT, the recovery time decreases as network connectivity increases.

\begin{figure}
\begin{center}
\includegraphics[scale=.45]{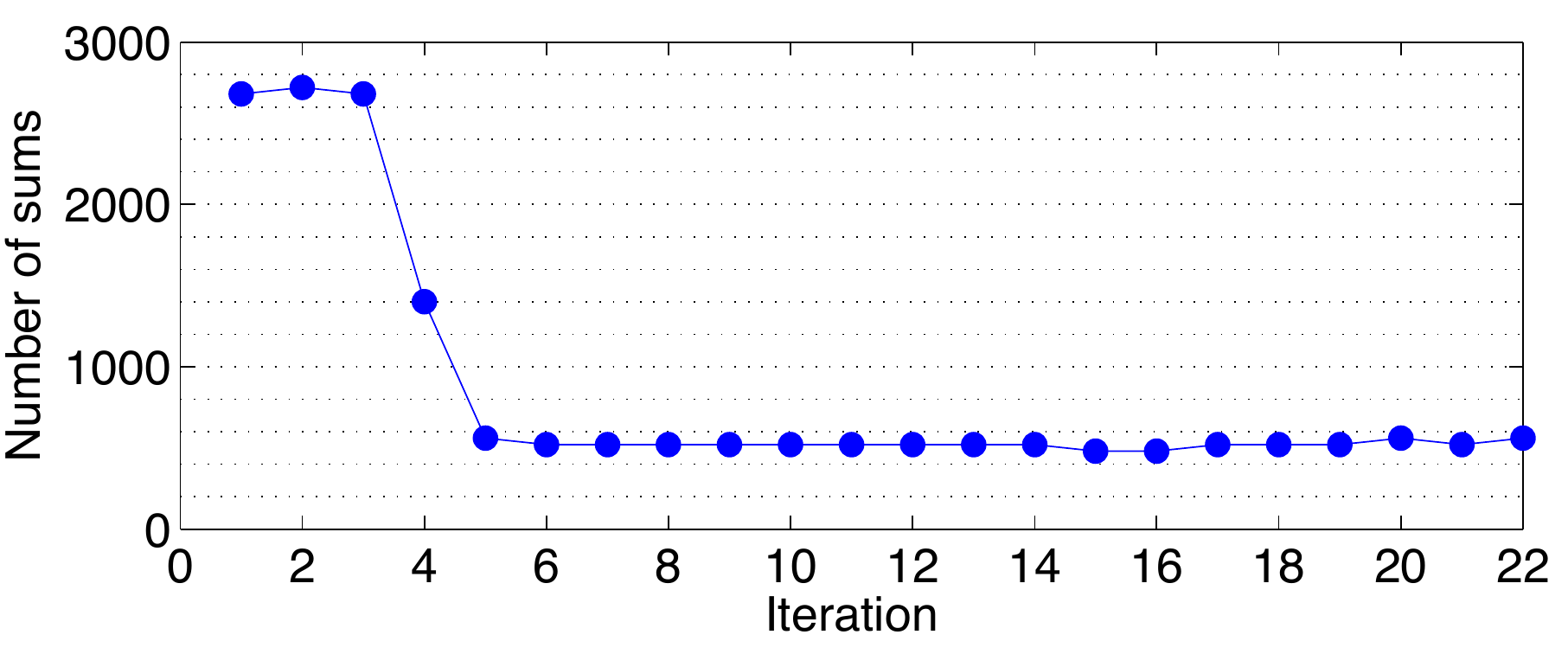}
\vspace{-.6cm}
\end{center}
\caption{Number of sums computed per iteration by D-IHT to solve Sparco problem 7 (with $N = 2560$) in a 40 node ER graph with connection probability of 0.25.}
\label{sums.fig}
\end{figure}

A key to the good performance of D-IHT is that, after just a few iterations, the algorithm finds the correct support set (the non-sparse components of the signal).
The magnitudes of the values in the support set quickly dominate the other values in the intermediate vectors.  As a result, in DATA, the sums for these objects are computed first, and the top-$K$ objects are identified after a minimal number of sum computations (on the order of $PK$).   
This behavior is illustrated in Figure \ref{sums.fig}, where we show the total number of sums computed for each iteration of D-IHT for a single experiment.  This figure shows a dramatic drop in the number of sum computations after just four iterations.

\bibliographystyle{IEEEbib}
\bibliography{cs}

\vfill \pagebreak

\appendix

\section{Pseudocode for DATA}
\begin{algorithm}
\caption{Distributed Absolute Threshold Algorithm, as it is executed by each node $p$.}
\SetAlgoNoLine
\SetAlgoNoEnd
\DontPrintSemicolon
%\SetNoFillComment
\KwFunction({\funcTopK($L^p = \{(ndx,val)\}_{i=1}^N$)}){ 
$topKList \gets \emptyset$, $top \gets 1$,  $bottom \gets N$  \;
$\topThresh \gets \infty$, $\bottomThresh \gets \infty$ \;
$done \gets \FALSE$ \;
\BlankLine
\While{$done = \FALSE$}{ 
\If{$\topThresh>\bottomThresh$}{
	$oid \gets$ new object id from top of list \;
} \Else{
	$oid \gets$ new object id from bottom of list \;
}
$\funcSum(oid)$ \;
\If{$p = 1$}{
 $\funcThresh(oid)$ \;   
}
\BlankLine
Receive $(ndx^q, sum^q)$ for $q=1 \ldots P$ \textbf{and}
receive $threshold$ \;
\If{$\topThresh>\bottomThresh$}{
$\topThresh \gets threshold$ \;
}\Else{
$\bottomThresh \gets threshold$ \;
}
\BlankLine
\For{$q = 1~\KwTo~P$}{
\If{$|topKList| < K$}{
		Add $(ndx^q,sum^q)$ to $topKList$ \;
	}\ElseIf{$sum^q >$ \mbox{min abs. sum in} $topKList$}{
		Replace smallest magnitude element with $(ndx^q,sum^q)$ \;
	}
}
\If{min. abs. sum in $topKList \geq \max(\topThresh,\bottomThresh)$}{
	$done \gets \TRUE$ \;
}
}
$x \gets \genX$($topKList$) \;
return $x$ \;
}
\label{alg:topk}
\end{algorithm}

\end{document}